# Observation of superconductivity in silicene

Lan Chen, Baojie Feng and Kehui Wu*

*Institute of Physics, Chinese Academy of Science, Beijing 100190, China*

*Corresponding author: khwu@iphy.ac.cn

**Abstract: A possible superconducting gap, about 35 meV, was observed in silicene on Ag(111) substrate by scanning tunneling spectroscopy. The temperature-dependence measurement reveals a superconductor-metal transition and gives a critical temperature of about 35-40K. The possible mechanism of superconductivity in silicene is discussed.**

Silicon is the basis of current semiconductor industry regardless of emerging proposals that materials such as diamond, carbon nanotube or graphene may replace silicon as the next generation semiconductors. It is of great interest if any novel property could be found in this "old" material, since it will then be easy to be integrated into the current semiconductor technology. For example, if silicon can be superconducting at reasonably high temperature, an era of "superconducting semiconductor industry" may immediately be foreseeable. Indeed, the discovery of superconductivity in boron doped diamond around 4 K [1], and $MgB_2$ at 39 K [2] had stimulated great interest in the superconductivity in silicon-based materials such as silicon carbide [3] or barium-doped silicon clathrates [4]. And soon it was found that highly boron-doped silicon is superconducting with the transition temperature ($T_c$) of 0.5 K [5] that is, however, too low to be of practical use. Another route for increasing $T_c$ is to find the materials with strong electron-phonon coupling contributing to the formation of Cooper pairs [6]. It was theoretically predicted that, in the 2D matierals- high doped graphane, the $T_c$ can be higher than boiling point of liquid nitrogen [7]. In this work we explored silicene, a single sheet of silicon atoms arranged in a honeycomb structure analogous to graphene [8]. It has attracted recent attentions because of the existence of Dirac fermion [9], stronger spin-orbital coupling than graphene [10], and compatibility with silicon-based nanotechnology. The physical properties such as the Dirac fermion behavior of charge carrier in silicene on Ag(111) surface has been experimentally conformed recently [11, 12].

In this Letter, we report the observation of a significant conductivity gap (Δ=35mV) in monolayer silicene on Ag(111) surface at temperature below 40 K. Several features, such as precise location of the peak at the Fermi energy ($E_F$), the DOS shoulders, indication of Andreev reflection, and gradual disappearance of the peak at temperature above 40 K, strongly suggest that this gap is a superconducting gap. The unusual high-$T_c$ (about 40 K) superconductivity in silicene might result from the strong electron-phonon coupling (EPC) and significant charge transfer from the Ag(111) substrate to silicene.

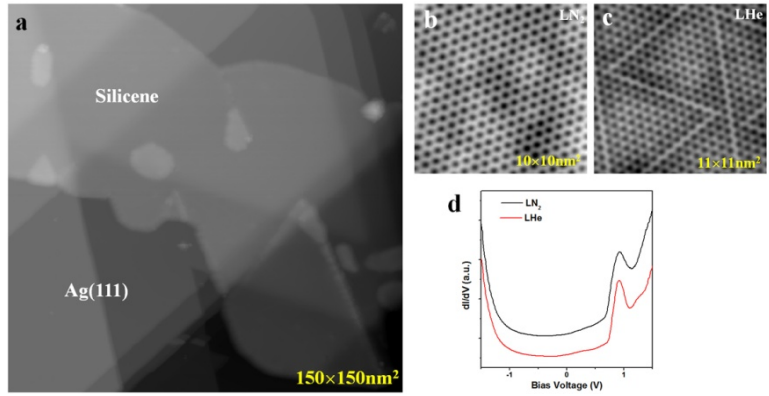

Fig.1 (a) STM image of a monolayer silicene film on Ag(111) substrate, running across several Ag(111) steps. (b) and (c) High resolution STM images of silicene taken at tip bias 1.0V at 77 K and 5 K, respectively. (d) dI/dV curves taken at 77 K (black curve) and 5 K (red curve), showing overall similar behavior (except the zero-bias gap which is too small to show in this voltage range).

The experimental condition and sample preparation procedure were identical to that in Ref.12. Fig.1(a) is an STM image showing a single layer silicene film on Ag(111) substrate, running across several steps of Ag(111) without losing the continuity. As we described in Ref. 12, at liquid $N_2$ temperature (77 K), monolayer silicene exhibits a honeycomb structure with a period of 0.64 nm (Fig. 1(b)), corresponding to (√3×√3)R30° superstructure with respect to the theoretical 1×1 silicene lattice. When cooled to liquid He temperature (5 K), silicene undergoes a phase transition, in which one of the two protrusions in each honeycomb unit cell becomes brighter than the other, showing a rhombic (√3×√3)R30° superstructure (Fig. 1(c)). As there are two possible configurations, the surface is phase separated into triangular domains with either one of the two symmetric configurations, separated by narrow domain boundaries where the neighbor protrusions are identically bright (Fig. 1c). The phase transition can be described by a "super-buckeling model" [13]. It has been known that free-standing silicene maintains a non-planar, so-called low-buckled (LB) geometry. While silicene is adsorbed on Ag(111) surface, this low buckled structure further adopts two mirror-symmetric (√3×√3)R30° rhombic super-buckled structures. The low transition barrier between these two phases enables dynamic flip-flop motion at high temperature, resulting in the (√3×√3)R30° honeycomb structure observed in STM [13].

Scanning tunneling spectroscopy (STS) probes the local density of states (LDOS). Typical dI/dV curves at wide energy range (from -1.5V to + 1.5V) obtained at 77 K and 5 K (Fig. 1(d)) reveal similar electronic structures, both with a small dip located at about 0.5 V attributed to the position of Dirac point (DP) of silicene, and a pronounced peak at 0.9 V. The similar electronic structures of the two phases confirm that the basic structures of the two phases are identical.

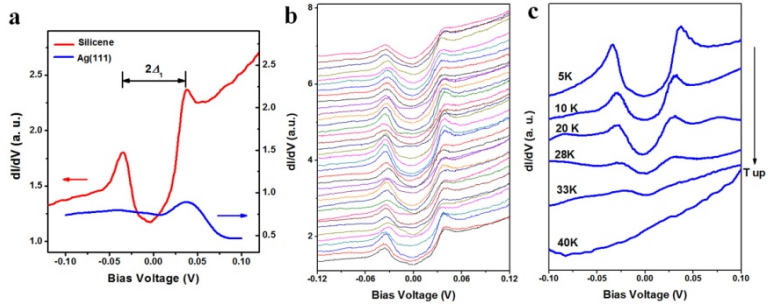

Fig.2(a) dI/dV curves obtained on silicene (red curve) and Ag(111) substrate (blue curve), respectively. (b) A series of dI/dV curves taken along a 15 nm line on silicene film, illustrating the uniformity of the gap. (c) Temperature dependence of dI/dV curve at same silicene film.

The dI/dV curve at narrow energy range around Fermi energy ($E_F$) (from -120mV to 120mV) is substantially different at 5K, as seen in Fig. 2(a). The spectra obtained on silicene shows a characteristic gap centered exactly at $E_F$ and two significant shoulder peaks at both sides. While the spectra taken on Ag(111) shows typical feature of LDOS of Ag(111), without any gap signature at $E_F$. Thus the gap is not induced by the Ag(111) substrate or STM tip. These observations have been reproduced on many different monolayer silicene films, with different tips. We also measured STS spectra on different locations on the silicene surface, and found that the gap exhibits high spatial homogeneity, as shown in Fig. 2(b).

The phenomenon that an energy gap opens at $E_F$ is usually explained by following mechanisms: (I) phonon-mediated inelastic tunneling or Kondo effect; (II) Peierls transition; (III) superconductivity. The gap induced by phonon-mediated inelastic tunneling in graphene can be as large as one hundred meV, but there are no DOS peaks at both sides of gap [14]. Similarly, there are also no peaks at the both side of the gap induced by Kondo screening in Kondo lattice such as $O_2$ monolayer on Au(110) surface [15]. So the mechanism (I) is ruled out. Secondly, if the structural phase transition of silicene observed in our experiments (Fig. 1(b) and (c)) is a Peierls transition [16], a gap should open at the DP of silicene, not around $E_F$ (The DP is at 0.5 eV below $E_F$ due to the charge transfer from Ag(111) to silicene). Moreover, we have illuminated that the phase transition in silicene is a dynamic phase transition, and not a Peierls transition [13]. Therefore mechanism (II) is also ruled out.

At last we believe that the gap should be a superconducting gap, and the two peaks are coherence peaks. We noted the intensity of LDOS in the gap region does not go to absolute zero, which might results from that the finite sample temperature (5K) in our experiments, which is not low enough. Note that the temperature sensor in our STM system is a bit far from the sample, so the error of sample temperature measurement can be about several Kelvin.

To confirm the observed superconductivity, we explored the superconductor-metal transition by varying the sample temperature. A sequence of dI/dV curves measured over the same monolayer silicene island at different temperatures shows that the coherence peaks can be clearly observed up to 28 K. When the temperature is increased to 33 K, the superconducting gap is still observable, but the two coherence peaks start to disappear. The gap is no longer visible at 40 K, and the curve reveals the metallic behavior without any gap. Such critical temperature (35 K~40 K) is the highest among all other single-element superconductors discovered so far, no matter it is 2D or 3D material. The robust superconducting gap $\Delta_1$=35 mV, which is half of the energy between the two coherence peaks, is much larger than conventional BCS superconductor, such as Nb [17], NbSe$_2$ [18] and even MgB$_2$ [19], and comparable with high-temperature superconductors such as cuprate superconductors [20]. Considering that Tc is smaller than 40 K in this case, the 2$\Delta_1$/kTc value is ~20 for this system, which is much larger than the BCS value of 3.52. It might mean that silicene on Ag(111) is not a conventional BCS superconductor. The dI/dV around the center of the gap (near E$_F$) exhibits V shape rather than U shape, which implies either the sample temperature is not low enough, or it does not have an *s* wave pairing symmetry.

The honeycomb structure of silicene is similar as that in B layer in MgB$_2$ and graphane. The high T$_c$ of conventional superconductor MgB$_2$ mainly stems from the strong electron-phonon coupling and the large charge transfer from Mg atoms to B layer [21]. The same mechanism has been applied to propose a superconductor of doped graphane with a T$_c$ above boiling point of liquid nitrogen [7]. In the case of silicene, the sp$^3$ hybrid electronic states have $\sigma$ characteristic. These $\sigma$ states are localized in the middle of the Si-Si bonds, and they should couple considerably to bond-stretching phonons. Furthermore, the buckling degree of Si atoms in ($\sqrt{3}\times\sqrt{3}$)R30° phase on Ag(111) (1.2Å higher than lower layer of Si atoms [13]) is much higher than LB model of free standing silicene (0.4 Å) [9], which may increase the $\sigma$ character in Si-Si bonds and result in the stronger electron-phonon coupling. The fact that the DP of silicene is at 0.5 eV below E$_F$ reveals significant charge transfer from Ag(111) surface to silicene, which results in large density of electronic states at E$_F$. These two factors may cooperate to establish high temperature superconducting states in silicene, with T$_C\approx$35-40K. Although silicene on Ag(111) may not be a conventional BCS superconductor, electron-phonon interaction may still play an important role in electron pairing.

We have performed STS measurements at different tip heights (smaller tunneling resistance means that the tip is closer to the surface), as shown in Fig. 3. When the tip approaches to surface, the tunneling conductance in the middle of the gap region is lifted, consistent with expected Andreev reflection [22] in this superconductor-metal junction. This provides another evidence of superconductivity in silicene. Another finding is the emergence of a smaller gap around E$_F$ with $\Delta_2$=15mV when the tip gets close to the surface, which implies that silicene on Ag(111) surface may be a two-band superconductor, similar as $MgB_2$ [21]. We note that, however, these features still need further confirmation, due to the finite cooling temperature and thus limited energy resolution in our system.

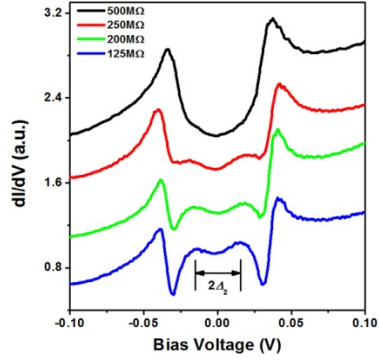

Fig. 3 dI/dV curves taken at different tunneling resistance of tunneling junction formed between STM tip and surface.

In summary, we have observed possible superconductivity in silicene on Ag(111) surface, at a temperature as high as 35-40 K, which is higher than all other single-element superconductors discovered so far. Though we believe the strong electron-phonon coupling in silicene play an important role in the formation of superconducting phase, the fundamental understanding of superconductivity such as pairing mechanism and pairing symmetry is still unknown and need to be investigated. A transport measurement is currently under investigation in order to provide a direct proof of the superconductivity in silicene, which is however challenging due to the oxidation of silicene in air and the high conductivity of the Ag substrate.

**Acknowledgements:** This work was financially supported by the MOST of China (Project No. 2012CB921703 and 2013CB921702), and the NSF of China (Project No. 11174344, No. 11074289, No. 91121003). L. Chen acknowledges the start funding from IOP, CAS.